# Use of Electronic Resources by Law Academics in India

Dr. Sunita D. Mane (Saware)
*PES's Modern Law College*, sunitamanesaware@gmail.com

Subaveerapandiyan A
*DMI-St. Eugene University*, subaveerapandiyan@gmail.com

# Use of Electronic Resources by Law Academics in India

Dr. Sunita D. Mane (Saware) and Subaveerapandiyan A
Librarian, PES's Modern Law College, Ganeshkhind, Pune
Librarian, DMI-St. Eugene University, Lusaka, Zambia
E-mail: subaveerapandiyan@gmail.com

**ABSTRACT**

This study investigated e-resources use, storage, the preferred format for reading, and difficulties faced while accessing e-resources. Electronic resources are playing a crucial role all over the world, and they are increasing widely in all age groups of the academic community. The main aim of the law academics' role is to know the effective use of electronic resources. For this study, we adopted a descriptive survey research design that was used to collect feedback from the respondents through the survey and Google form. The study samples are Progressive Education Society's Modern Law College affiliated with Savitribai Phule Pune University. BA LLB students are samples of the study. The study findings reveal that 95.1% of student respondents are aware of e-resources; 75.7% of respondents used college websites for accessing legal electronic resources. The most noticeable is that 95.1% of users prefer reading in Portable Document Format (PDF). Overall, more than 50% of respondents felt e-resources are easy to access, time-saving, and search tools. The findings of this study are helpful to library professionals to subscribe to more e-resources. The research suggests that library professionals have to guide the users to conduct orientation on accessing the e-resources more effectively. Students expected user manuals for accessing some databases and e-resources based on the findings, recommendations were made.

## 1. INTRODUCTION

Most legal academics preferred e-resources to e-journals, e-judgments, e-books, and legal databases in the twenty-first century because of their broad breadth, extensive indexing and analytical content, and rapid updates. These high-end databases are crucial in the legal profession, where "time is money", and expert analysis informs legal counsel. Law students must be adept users of these databases to compete for and succeed in most law careers. Print resources and less advanced internet technologies are available to aid scholars. It is impossible to overestimate the value of library resources in legal education. Nowadays e-resources are becoming more popular since they are not only easy to use but are also reliable. Searching for relevant information in a time frame is significant for both researchers and information professionals. E-resources are gaining popularity due to their easy and fast accessibility. The new and current trend of library services has an economic and user-oriented tendency to access library collections electronically. Library professionals feel that electronic information has a vital role in providing updated information and timely delivery to the users.

Graduate students, according to Wu and Chen (2012)[1], are frequent users of electronic library resources and believe that they are crucial to their studies. According to Tenopir (2003)[2], are "binge" consumers of e-journals, searching for journal articles regularly during their thesis-writing phase. Law students have a vast spectrum of reference requirements, frequently linked to upper-level coursework, such as seminar papers and extracurricular activities, such as journal participation and externships. A rudimentary introduction to legal materials and search strategies are also required for first-year law students. E-resources are suited for improving library services in various ways, according to the study's goals. To begin with, most e-resources include sophisticated search and retrieval facilities that help law students conduct more successful and efficient literature searches. Modern Law College has a wealth of e-resources available to law students via the college's websites. The research on electronic resources and the level of student satisfaction are optimistic.

### 1.1 ELECTRONIC RESOURCES

Electronic resources are sources that provide on-time information in an electronic format, and the information is available at any time as per the need of users. Electronic resources are enabled by a technical capability to create a search by using an enormous amount of information. Electronic resources include electronic journals, electronic books, electronic databases, CD ROMs, DVDs, Internet resources, etc. The characteristics of E-Resources are multi-access, speed, functionality, content reuse, management, storage timeliness etc. Information access can be made without wasting any time coping with open mouth problems of space and budget etc. in libraries and information centres desired information can be retrieved within a few minutes at the learning desk vast collection of information can be stored in a small place Resource sharing at the desired level among libraries and information centres.

**Definitions:** According to AACR2, 2005 Update, an electronic resource is: "Material (data and/or program(s)) encoded for manipulation by a computerized device. This material may require the use of a peripheral directly connected to a computerized device (e.g., CD-ROM drive) or a connection to a computer network (e.g., the Internet)." This definition does not include electronic resources that do not require the use of a computer, for example, music compact discs and video disks.

According to the Library and Information Technology Glossary "Term used to describe all of the information products that a library provides through a computer network... .." According to Wikipedia, Electronic Resources means "Information (usually a file) which can be stored in the form of electrical signals, usually on a computer; Information available on the Internet"[3].

### 2. ABOUT PROGRESSIVE EDUCATION SOCIETY MODERN LAW COLLEGE

Progressive Education Society's Modern Law College was established in 2003 in Pune, Maharashtra, affiliated with one of the oldest and most renowned universities, Savitribai Phule Pune University. PES Modern Law College is one of the leading legal and educational institutions. It was accredited B++ grade by National Assessment Accreditation Council (NAAC) and recognized by the Bar Council of India. College offers various law courses such as B.A. LL. B (5 years), B.B.A.LL. B (5 years), LL.B. (3 years), Masters in Laws (2 years), PhD in Law, along with various Diploma programs and Certificate programs. The college provides various library services and subscribes to multiple e-resources and e-databases. The library has 8893 print books, 42 subscribed journals, reports, and E-information resources. The library facilities and services are reprographic, scanning, Internet, OPAC, newspaper clipping, indexing, references services, and many more[4-5].

### 3. OBJECTIVES OF THE STUDY:

This research aims to look into the use of electronic resources by legal experts at Modern Law College, Savitribai Phule Pune University, Pune.

- ➢ To know the use of electronic resources
- ➢ To identify the preference for electronic format for reading
- ➢ To examine the problems faced by the users while accessing and using electronic resources.
- ➢ To understand the awareness of users about available electronic resources
- ➢ To study the satisfaction level of users to access the electronic resources

### 4. REVIEW OF LITERATURE

Literature reviews are a collection of the most relevant and extensive publications on e-resources to comprehensively express various researchers. Many researchers focused their efforts on researching electronic resources. Ten years ago, reviews of the literature were mentioned, where it was noticed that we conducted extensive research studies in India and abroad.

Deng (2010)[6] used an online survey to study emerging patterns and trends in using electronic resources in a higher education environment in Australia. They found 305 valid responses out of a total student population of about 57000. Because of the survey's inadequate information on the size of the actual population, calculating the response rate is challenging. Most respondents thought electronic resources were beneficial. They also discovered that users with different purposes accessed and used electronic resources in significantly different ways. The availability and quality of information are essential factors that affect electronic resources.

Bhatt and Singh Rana (2013)[7] conducted a study on the use of e-resources by the engineering academics of Rajasthan state. As per the aim of this study, over 65% of professors/readers and lecturers are given the top priority behind the use of e-resources. Maximum users were quite satisfied with using e-resources because of their technical or hardware/software problems, low-speed connectivity, high cost, doubts about the permanence of e-journals and ebooks archive, etc., again 60% of users were not aware of the statutory provision for providing e-resources by their respective institutions.

The Impact of Electronic Resources in Portuguese Academic Libraries: Results of a Qualitative Survey was undertaken by Melo and Pires (2010)[8]. They collected the data using an e-mail survey sent to thirty-three Portuguese public universities. They chose the performance indicators to be evaluated using the International Standards ISO 11620:1998, Amendment 1:2003 new performance indicators for libraries, and ISO 2789:2006. As per the evaluation, they received 1786 responses in two months. Professor/Researcher 32.4%, PhD. Students 7.8%, Master and Undergraduate Students 27.8%, and Other 32.0% (administrative and library staff) per cent made up the population. They found the contingent valuation approach, utilizing the estimated value of the time saved to measure the benefits.

Sohail and Ahmad (2017)[9] discuss using electronic resources and services by Fiji National University faculty and students. Libraries' e-resources and services are critical to the operation of any academic organization and nation-building efforts. These e-resources and services are frequently accessed. According to the findings of this study, most respondents are aware of the search options for accessing e-resources.

Ankrah and Atuase (2018)[10] Postgraduate students at the University of Cape Coast use electronic resources. The researcher described the e-resources and observed that postgraduate students did not use their e-resources due to low publicity, inadequate training, access restrictions such as passwords and usernames, and other limitations such as poor internet connection, inadequate computers, power outage, and inadequate searching skills, which forced students to rely more on library professionals for their information searches.

According to Gul and Bano[11] smart libraries are the next generation of libraries that work with a combined effect of smart technologies, smart users, and smart services. The study confirms that smart libraries are becoming smarter as new smart technologies emerge, which improves their working capabilities and satisfies the users who use them. Using smart technologies in libraries has helped bridge the gap between the services provided by libraries and the rapidly changing and competing needs of humans.

Santhi (2020)[12] researched the 'Use of Electronic Resources in Indian Academic Institutions in India. The researcher briefly discusses various published and unpublished literature on various topics such as the impact of e-journals, the impact of online journals, library consortiums, the use of e-journals, usability, user attitude toward e-journals, where information and communication technology (ICT) policies differ in the social organization, the significant four limitations are reflected, studies are frequently restricted to a single discipline, with a focus on useful resources rather than non-use, which skews patterns observed within and across disciplines.

Leonard et al. (2020)[13] present a case study of a Namibian university. They discuss the level of awareness of electronic resources available to them, how beneficial and effective they found these e-resources—the difficulties they face in retrieving them. Law lecturers' challenges in using e-resources are aware of and use e-resources for lesson preparation and research. There is an urgent need to advance electronic law collection, Internet bandwidth, promotion, and training frequency on how to access and successfully search for electronic information.

Habiba et al (2022)[14] published an article titled 'Information Behaviour of Faculty Members of NSTU, Public University of Bangladesh'. The primary goal of this study was to look into the behaviours of faculty members while conducting research. The research showed that faculty members are dependent on search engines to access information. They primarily used academic social media sites such as Google Scholar, Research Gate, and discussion lists with new publications. In their search techniques, they discovered federated tools and Boolean operators.

## 5. METHODOLOGY

### 5.1 Research design
The study was guided by five objectives formulated and a descriptive survey research design was also adopted. According to Quadri & Idowu (2016), [15] methods rely heavily on quantitative methods of data gathering and acknowledged that the survey method is popular and mostly utilized in the field of humanities as well as social sciences.

### 5.2 Populations
We conducted the study with PES's Modern Law College affiliated with Savitribai Phule Pune University. The study sample comprises Bachelor of Legislative Law Students. The questionnaire was the main instrument used for data collection, it contained 12 items and was prepared and validated by Google Spreadsheet. We collected the data with the help of Google Forms. For this study, quantitative and qualitative methods were used, and survey questionnaires were adopted. We used descriptive statistics frequency and the percentage calculated. We collected the data from January 13, 2022, to Jan 31, 2022. The researchers were studying the use of electronic resources among the students of Modern Law College; therefore, descriptive research is the appropriate method for the study undertaken for the research by the researchers. We distributed questionnaires to 448 Bachelor of Legislative Law (BA LLB) five year integrated undergraduate students of Modern Law College affiliated with Savitribai Phule Pune University. We received 226 questionnaires from students, 119 males and 107 females. The present study was concerned with using electronic resources by law academics.

## 6. DATA ANALYSIS AND FINDINGS

Data were analyzed with the use of descriptive and inferential statistics. Research questions were analyzed using percentage counts and respondent tables.

**Table 1: Do you have an awareness of e-resources?**

| Awareness of e-resources | Frequency | Percentage |
|---|---|---|
| Yes | 215 | 95.1 |
| No | 11 | 4.9 |

Table 1 shows information regarding the students' awareness of e-resources at Modern Law College. The result reveals that most of the students, 215 (95.1%), are aware of e-resources, and very few 11 (4.9%) students have less awareness. Thus, based on the study results, it is evident that the majority of the law students of Modern Law College have awareness of e-resources.

**Table 2: Have you visited your College Library website to access legal e-resources?**

| Accessing legal electronic resources | Frequency | Percentage |
|---|---|---|
| Yes | 171 | 75.7 |
| No | 55 | 24.3 |

Table 2 reveals that 171 (75.7%) students visited the college library websites to access the legal resources, and 55 (24.3%) of students did not visit their library websites. It is observed from the collected data, that most of the students have access to legal e-resources directly through the college library website.

**Table 3: If yes, which categories of electronic resources are used?**

| Electronic resources used | Frequency | Percentage (N=226) |
|---|---|---|
| E-Journals/Magazines | 67 | 29.6 |
| E-Books | 87 | 38.5 |
| Legal Databases | 74 | 32.7 |
| E-Thesis and Dissertation | 26 | 11.5 |
| E-Newspapers | 51 | 22.6 |
| Law-related useful links | 100 | 44.2 |
| Syllabus | 142 | 62.8 |
| Subject PPT | 87 | 38.5 |
| Encyclopedias & Current News | 32 | 14.2 |
| Govt. Organization Websites | 55 | 24.3 |
| E-Judgements | 46 | 20.4 |
| Assignment questions | 60 | 26.5 |
| SPPU Digital Repository (Law Study Materials) | 57 | 25.2 |
| Open Access Law Journals | 38 | 16.8 |

Table 3 provides the details of electronic resources used by undergraduate law students. The majority of the respondents, 142 (62.8%), use the syllabus. 100 (44.2%) used law-related useful links. E-books and subject PPTs were used by 87 (38.5%) of respondents. 74 (32.7%) of respondents used legal databases. 67 (29.6%) of respondents used e-journals/magazines. 60 (26.5%) of respondents used assignment questions. Followed by one-quarter of respondents, 57 (25.2%) used SPPU digital repository (law study materials). 55 (24.3%) of respondents used Government organization websites. (22.6%) of respondents used E-newspapers. 46 (20.4%) of respondents used e-judgements. All other e-resources used by less than 20% of respondents, such as open access law journals, Encyclopedias & current news and e-thesis and dissertation. Thus, based on the study results, indicates that they have to prefer law-related study materials to useful links.

**Table 4: Which format do you usually use for reading?**

| The format usually uses for reading | Frequency | Percentage (N=226) |
|---|---|---|
| PDF | 215 | 95.1 |
| Word | 86 | 38.1 |
| PPT | 116 | 51.3 |
| Excel | 27 | 11.9 |
| HTML | 24 | 10.6 |

Table 4 represents students' favourite format of e-resources for reading. The highest 215 (95.1%) of respondents' choices are in pdf format. 116 (51.3%) of respondents like PPTs. 86 (38.1%) of respondents like Word format. 27 (11.9%) respondents liked the Spreadsheet format, and 24 (10.6%) respondents used the HTML format. E-resources are available in various multimedia formats even though most users prefer PDF(Portable Document Format).

**Table 5: Which factors are effective to access electronic resources?**

| Factors are effective to access electronic resources | Frequency | Percentage (N=226) |
|---|---|---|
| Easy to access | 177 | 78.3 |
| Time-Saving | 159 | 70.4 |
| Availability of Search Tools | 118 | 52.2 |
| Variety of Resources | 103 | 45.6 |
| No Physical Space Limitations | 68 | 30.1 |

The respondents were asked effectively to access e-resources. 177 (78.3%) respondents emphasized it is easy to access, and 159 (70.4%) respondents pointed out it saves time. 118 (52.2%) of respondents chose the availability of search tools. 103 (45.6%) respondents say a variety of resources we can use. 68 (30.1%) of respondents mentioned no physical space limitation.

**Table 6: Where do you store your e-resources?**

| Storage of E-resources | Frequency | Percentage (N=226) |
|---|---|---|
| Cloud Storage | 151 | 66.8 |
| Physical Storage | 87 | 38.5 |
| Internal Hard Disk Drive | 70 | 31 |
| Hard Copy (Printout) | 67 | 29.6 |

Table 6 exhibits e-resources shared location by the users. 151 (66.8%) of respondents were stored by clouds such as Google Drive, Dropbox, Mega and Cloud, etc. 87 (38.5%) used physical storage devices external, e.g. External Hard Disk, Pen Drive, etc. 70 (31%) used internal hard disk drives and 67 (29.6%) used hard copy (printouts).

**Table 7: What type of difficulties do you have to face while accessing electronic sources?**

| Challenges faced by you while accessing e-resources | Frequency | Percentage (N=226) |
|---|---|---|
| Lack of information about how to use e-resources | 60 | 26.5 |
| Lack of time for searching | 63 | 27.9 |
| Information is scattered in too many sources, so difficult to explore without the assistance | 84 | 37.2 |
| Non-availability of needed e-resources | 47 | 20.8 |
| Lack of technical support | 49 | 21.7 |
| Lack of training to use the e-resources | 26 | 11.5 |
| Lack of support from the library | 21 | 9.3 |
| I did not feel any difficulties | 8 | 3.5 |

Data presented in table 7 is that of problems based on students while accessing e-resources. 84 (37.2), the more respondents expect assistance for searching the information because the information is scattered in plenty of sources. 63 (27.9%) respondents feel a lack of time exploring the e-resources. 60 (26.5%) of respondents mentioned that user guides are not available to access and use the information resources. 49 (21.7%) responses highlighted a lack of technical support. 47 (20.8%) of respondents felt that non-availability of needed e-resources (e-journals and e-databases). 26 (11.5%) lack training to use the e-resources/products. 21 (9.3%) mentioned a lack of support from the library. The least respondents, 8 (3.5%), did not feel any difficulties.

**Table 8: What is your opinion about e-resources?**

| Opinion about e-resources | Frequency | Percentage (N=226) |
|---|---|---|
| I can do better research because of the availability of e-resources | 169 | 74.8 |
| Some of the research information needed is now only available online | 71 | 31.4 |
| More challenging to find required information while using e-resources | 39 | 17.3 |
| Still, I prefer access to print as well as e-resources | 63 | 27.9 |
| More comprehensive information is available | 54 | 23.9 |

Data presented in table 8 indicated students' opinions about e-resources and their usage. More than 169 (74.8%) of respondents agreed that the availability of e-resources can help for better research. The most minor 39 (17.3%) respondents only felt it was challenging to find needed information while using e-resources.

## 8. DISCUSSION OF FINDINGS

From the above data of Tables and Graphs, it is found that the following findings can be concluded which are as follows:

- ➢ This study showed that the majority of the students of Modern Law College are aware of e-resources.
- ➢ The syllabus is one of the electronic resources used by undergraduate law students. 44.2% of students used law-related informative links from college websites. 74.8% of students have to use e-books, subject PPTs, and legal databases to learn new things, and 54.4% for their research.
- ➢ As per the study, 37% of students face difficulties accessing the e-resources because of non-availability of assistance, lack of technical support, lack of time, and lack of support from the library.
- ➢ User training is essential to improve the facilities and services for effective use of e-resources.
- ➢ More funds are also vital to subscribe to more e-resources. A library technical assistant full-time post is also necessary to provide better services to the users.

## 9. CONCLUSION

Modern Law College students can benefit immensely from the enormous usage of e-resources, particularly for teaching, learning, and research. It must be noted that students who maximally utilize the available e-resources will no doubt excel academically. However, it was discovered from the findings that law students are aware of the e-resources and use them for various purposes ranging from assignment completion, research work, and project work as well as to updating knowledge. Also, the e-resources were used daily as an online search engine, computer and internet facilities emerged, the daily used e-resources while e-book, e-journal, and CD ROM databases were used occasionally. However, some of them faced challenges among the students; support of the subordinators regarding the e-resources is not found at the optimum level, so the library must take initiatives to organize orientation programs and user awareness programs.

## 10. RECOMMENDATIONS

Based on the findings, the following recommendations were made:

(a) Modern Law College management should think about developing a functional policy to enhance the efficient use of e-resources among the students. Also, it is recommended that the policy should focus on individual intellectual property as well as technological infrastructure.

(b) Modern Law College students should also engage in periodical training, particularly on emerging technology that will assist them in using the available e-resources.

(c) College should make funds available for law studies to subscribe to current and relevant e-resources that will foster students' academic performance.

(d) The need for more qualified staff with better exposure to modern technological devices to assist users when they face problems in accessing e-journals should be stressed.